\documentclass[conference]{IEEEtran}
\IEEEoverridecommandlockouts
\usepackage{cite}
\usepackage{amsmath,amssymb,amsfonts}
\usepackage{algorithmic}
\usepackage{graphicx}
\usepackage{textcomp}
\usepackage{xcolor}

\def\BibTeX{{\rm B\kern-.05em{\sc i\kern-.025em b}\kern-.08em
    T\kern-.1667em\lower.7ex\hbox{E}\kern-.125emX}}

\usepackage{siunitx}
\usepackage{lipsum} 
\usepackage{comment}
\usepackage{balance}
\usepackage{amssymb}
\newcommand{\C}{\mathbb{C}}
\usepackage{xspace}

\newcommand{\name}{AESPA\xspace}
\usepackage{enumitem} 
\usepackage{multirow} 
\usepackage{makecell}

\begin{document}

\title{\name: Accuracy Preserving Low-degree Polynomial Activation for Fast Private Inference\\}
\author{\IEEEauthorblockN{Jaiyoung Park}
\IEEEauthorblockA{\textit{Seoul National University}\\
jypark@scale.snu.ac.kr
}
\and
\IEEEauthorblockN{Michael Jaemin Kim}
\IEEEauthorblockA{\textit{Seoul National University}\\
michael604@scale.snu.ac.kr
}
\and
\IEEEauthorblockN{Wonkyung Jung}
\IEEEauthorblockA{\textit{Seoul National University}\\
wkjung@scale.snu.ac.kr
}
\and
\IEEEauthorblockN{{Jung Ho} Ahn}
\IEEEauthorblockA{\textit{Seoul National University}\\
gajh@snu.ac.kr}
}

\maketitle
\thispagestyle{plain}
\pagestyle{plain}

\begin{abstract}
Hybrid private inference (PI) protocol, which synergistically utilizes both multi-party computation (MPC) and homomorphic encryption, is one of the most prominent techniques for PI.
However, even the state-of-the-art PI protocols are bottlenecked by the non-linear layers, especially the activation functions. 
Although a standard non-linear activation function can generate higher model accuracy, it must be processed via a costly garbled-circuit MPC primitive.
A polynomial activation can be processed via Beaver's multiplication triples MPC primitive but has been incurring severe accuracy drops so far.

In this paper, we propose an accuracy preserving low-degree polynomial activation function (\name) that exploits the Hermite expansion of the ReLU and basis-wise normalization.
We apply \name to popular ML models, such as VGGNet, ResNet, and pre-activation ResNet, to show classification accuracy comparable to those of the standard models with ReLU activation, achieving superior accuracy over prior low-degree polynomial studies. When applied to the all-ReLU baseline on the state-of-the-art Delphi PI protocol, \name shows up to 61.4$\times$ and 28.9$\times$ lower online latency and communication cost.
\end{abstract}


\section{Introduction}
\label{sec:introduction}

\emph{Private inference (PI) protocols}~\cite{liu_2017_oblivious,lehmkuhl_2020_delphi,juvekar_2018_gazelle}, or hybrid PI protocols exploiting multi-party-computation (MPC~\cite{yao_1982_protocols}) and homomorphic encryption (HE~\cite{gentry_2009_fully}), are one of the most prominent approaches in machine-learning-as-a-service (MLaaS) to support sensitive data, such as medical images~\cite{agrawal_2019_quotient, riazi_2019_xonn}.
In these protocols, MPC primitives and HE are combined in a complementary manner (e.g., HE is used for linear layers and MPC primitives used for non-linear layers).
Although promising, the performance (i.e., serving latency) of the PI protocols is commonly bottlenecked by the non-linear layers of the machine learning (ML) models, especially activation functions~\cite{juvekar_2018_gazelle, lehmkuhl_2020_delphi}.
While standard activation functions such as ReLU can generate higher model accuracy, they rely on slow and communication/storage-hungry primitives such as garbled-circuit (GC) MPC primitive~\cite{yao_1986_how} due to the required bitwise operations.
In contrast, the polynomial activation functions have an advantage in that they can be processed via less costly Beaver’s multiplication triples (BT) MPC primitive~\cite{beaver_1995_precomputing}, but are limited by the lower model accuracy. 

Prior works have tackled this challenge in various ways. 
First, one set of prior arts sought the effective approximations of the standard activation functions, such as ReLU, which minimize the model accuracy degradation and evaluation cost at the same time. \cite{gilad_2016_cryptonets,garimella_2021_sisyphus,ishiyama_2020_highly,obla_2020_effective,hesamifard_2019_deep,thaine_2019_efficient,lou_2021_hemet,lee_2021_precise} explored various polynomial activations to replace ReLU, especially based on numerical analysis such as Taylor polynomials or minimax approximation~\cite{veidinger_1960_numerical}. Circa~\cite{ghodsi_2021_circa} broke down ReLU into a piecewise linear function and a sign function to partly utilize BT, and approximating sign for reducing the cost of GC.

However, the ML models with 
such approximated activation
often suffer from either limited trainability when the model is deep for a large task, low model accuracy, or high serving latency when pursuing high accuracy.
\cite{gilad_2016_cryptonets} was limited to a shallow model for a simple task of MNIST.
\cite{lee_2021_precise} minimized the accuracy drop for deeper models with larger tasks of CIFAR-10~\cite{krizhevsky_2009_cifar} and ImageNet~\cite{russakovsky_2015_imagenet} by exploiting the minimax approximation function.
However, it was limited from the long latency caused by using high polynomial degrees (e.g., 29) for approximation; a higher degree means that more MPC primitives must be used. Circa~\cite{ghodsi_2021_circa} demonstrated a better (lower) latency, but still leaves a large room for improvement.


\begin{figure}[tb!]
    \begin{center}
    \includegraphics[width=1\columnwidth]{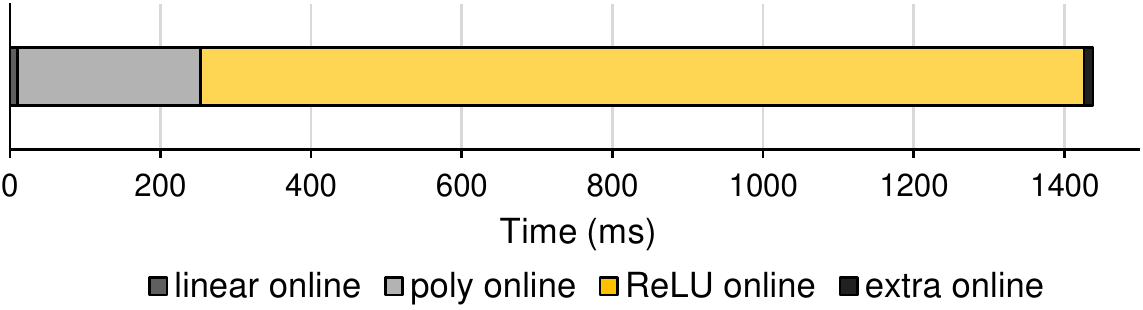}
    \end{center}
    \vspace{-0.2in}
    \caption{The online latency breakdown of Delphi~\cite{lehmkuhl_2020_delphi} for ResNet32, where 26 ReLU functions are
    approximated with quadratic polynomials and processed via Beaver's multiplication triples, 
    whereas 5 ReLUs are processed as via Garbled circuit. Experimental details in Section~\ref{sec:4_experimental_setup}.} 
    \label{fig:1_time_breakdown}
    \vspace{-0.2in}
\end{figure}

Second, another set of studies~\cite{lehmkuhl_2020_delphi,lou_2020_safenet} utilized the neural architecture search (NAS) method~\cite{zoph_2017_neural} to find optimal activation layer locations (in terms of model accuracy and online inference latency) to substitute with the polynomial activation while leaving others with the original ReLU. 
%
However, despite the use of NAS, the acquired models still exhibit large room for improvement in both accuracy and latency because not all activation layers can be replaced with polynomials, and the remaining few activation layers take up the majority of the total online latency. 
Delphi~\cite{lehmkuhl_2020_delphi}, a state-of-the-art PI protocol utilizing NAS, greatly reduced the online latency, achieving 3.8 seconds and 65.7\% accuracy for ResNet32 on CIFAR-100, when 26 out of 31 ReLUs are replaced into a quadratic approximation, for instance.
However, up to 82\% of the online latency is  spent on GC for the five remaining ReLUs (see Figure~\ref{fig:1_time_breakdown}); no more ReLUs can be replaced due to the sharp drop in the model accuracy.


In this paper, we propose an \textbf{A}ccuracy pr\textbf{ES}erving low-degree \textbf{P}olynomial \textbf{A}ctivation (\name) that can replace all ReLUs in the neural networks, using the \emph{Hermite polynomial} concatenated with the \emph{basis-wise normalization}.
While prior studies~\cite{lokhande_2020_generating, ge_2017_learning, agarwal_2021_deep, panigrahi_2020_effect} inspired our work, they were mainly in the theoretical domain.
None had explored nor demonstrated the practicality of the Hermite polynomials in the PI context or was able to reach high accuracy as our method. 
Our composition of Hermite expansion and basis-wise normalization substantially mitigated vanishing (or exploding) gradient problems, which are frequently encountered with the polynomial activation functions, allowing us to achieve high model accuracy even using the low-degree polynomials.

We first present the organization of our Hermite polynomial with the basis-wise normalization (HerPN) block and show how it can be used to modify ResNet~\cite{he_2016_deep}, pre-activation ResNet (PA-ResNet~\cite{he_2016_identity}), and VGG~\cite{simonyan_2014_very} (Section~\ref{sec:3_contribution}).
Then, we report the smaller (better) accuracy drop of the trained models using the HerPN blocks for CIFAR-10/100, and TinyImageNet datasets~\cite{yao_2015_tiny} over the prior polynomial activation ML models of \cite{lee_2021_precise,garimella_2021_sisyphus}.
Moreover, we demonstrate the effectiveness of the HerPN-modified ML models for a state-of-the-art PI protocol, Delphi, on the real machine and network (Section~\ref{sec:4_evaluation}).
By replacing all ReLUs with our HerPN blocks and processing them with BT, the online latency and communication for evaluating the ML models, such as ResNet18/32, PA-ResNet18/32, and VGG16, are reduced by up to 61.4$\times$ and 28.9$\times$, respectively.
The accuracies of the ML models with the HerPN block are also competitive with the original models with ReLU for CIFAR-10/100 and TinyImageNet.

The key contributions of this paper are as follows:
\setlist{nolistsep}
\begin{itemize}[noitemsep,leftmargin=0.2in]
  \item We propose a novel low-degree polynomial activation function that utilizes the Hermite polynomial concatenated with the basis-wise normalization, which sustains the high accuracy of the ML model.
  \item We demonstrate superior accuracy over prior polynomial activation works on large datasets such as CIFAR-10/100.
  \item We showcase the effectiveness of our work in the PI context, in terms of latency and communication costs, on a real machine and network using the open-source Delphi PI protocol.
\end{itemize}

\section{Background}
\label{sec:2_background}

\subsection{Private inference (PI) and hybrid PI protocols}
\label{sec:2_pi_protocols}

Private inference (PI) refers to a set of techniques that enable machine learning (ML) inference without the need to reveal i) the private data of a client to a service provider or ii) the trained model of a service provider to a client.
Fully-homomorphic-encryption (FHE)~\cite{chillotti_2020_tfhe, gentry_2009_fully, cheon_2017_homomorphic, fan_2012_somewhat}, federated learning~\cite{bonawitz_2019_towards, konecny_2016_federated}, or works based on enclaves~\cite{costan_2016_intel, ngabonziza_2016_trustzone} all pursue practical PI.
However, none has proven to be dominantly superior to the others regarding the security level, latency, or accuracy of the supporting ML models.
In particular, while FHE might be a promising option for the PI, it is limited by i) the extremely costly bootstrapping operation to enable an unlimited depth of operations for deep ML models and ii) the limited applicability to arbitrary operations.

Hybrid \emph{PI protocols} based on both MPC primitives and HE attempt to provide a prominent option for private inference~\cite{juvekar_2018_gazelle, lehmkuhl_2020_delphi,liu_2017_oblivious}.
These PI protocols often exploit the leveled-HE (LHE, which does not support bootstrapping) for ML \emph{linear layers}.
As LHE evaluates a single linear layer at a time, no bootstrapping operation is necessary, thus incurring a shorter computation time than FHE.
A PI protocol often processes the non-linear activation function via MPC primitives.
MPC primitives require minimal computation time at the cost of high communication costs between the client and the server.
Our work focuses on the latter of processing activation functions with the MPC primitives.

\subsection{Cryptographic primitives}
\label{sec:2_crypto_primitives}

The hybrid PI protocols rely on multiple cryptographic primitives, either the ones from MPC or HE, to effectively support the private inference. Parameters about the finite field and data representation in cryptographic primitives are explained in Appendix \ref{sec:appendix_b}.

\noindent
\textbf{Additive secret sharing} (SS~\cite{shamir_1979_share}): SS is a cryptographic primitive that divides a secret value $\langle x\rangle$ into multiple shares $\langle x\rangle_1,\langle x\rangle_2,...,\langle x\rangle_n$ distributed to $n$ parties so that a single party does not have complete knowledge of the original secret.
In a two-party SS, the secret shares of the value $x$ can be generated by randomly sampling a value $r$ and distributing the shares $\langle x\rangle_1\!=\!r$ and $\langle x\rangle_2\!=\!x\!-\!r$ to each party. 
The reconstruction of the original secret is straightforward, by pooling shared values from the two parties and adding up: $\langle x\rangle\!=\!\langle x\rangle_1\!+\!\langle x\rangle_2$.

\noindent
\textbf{Beaver’s multiplication triples} (BT~\cite{beaver_1995_precomputing}): BT is a two-party protocol that can securely compute the product of the two secret-shared values.
The whole protocol is divided into two steps: generation and multiplication procedures.
For the first step to prepare for the multiplication, Beaver’s triples are generated; randomly sampled $a$, $b$, and $a\!\cdot\!b$ are secret-shared between two parties $P_1$ and $P_2$.
As a second step, when the actual input $x$ and $y$ are already secret-shared (namely $P_1$ holds $\langle x\rangle_1, \langle y\rangle_1$ and $P_2$ holds $\langle x\rangle_2, \langle y\rangle_2$), through Beaver's multiplication procedure the SS of their product is generated and shared (namely $\langle xy\rangle_1$ for $P_1$ and $\langle xy\rangle_2$ for $P_2$).
During this second step, one set of Beaver’s triples is consumed for every multiplication.

\noindent
\textbf{Garbled circuit} (GC~\cite{yao_1986_how}): GC is a two-party protocol that enables 
garbler and evaluator to securely compute an arbitrary boolean circuit ($C$) without revealing their private inputs.
Initially, a garbler holds its private data $x_g$ and boolean circuit $C$, while an evaluator holds its private data $x_e$.
The protocol starts with the garbler encoding (garbling) the circuit to generate garbled circuit $\hat{C}$, which takes encoded inputs (labels) and evaluate the original circuit $C$.
The garbler sends $\hat{C}$ and encoded data $L(x_g)$ to the evaluator.
The label for the evaluator’s input, $L(x_e)$, can be computed by using the Oblivious Transfer protocol\footnote{Oblivious Transfer (OT) is also a type of MPC primitives. OT allows the sender to transfer only one of multiple possible data without knowing what has been transferred to the receiver.}~\cite{rabin_2005_exchange} with the garbler.
Finally, the evaluator computes $Eval(\hat{C}, L(x_g), L(x_e))$, which outputs  $y=C(x_g, x_e)$.

\noindent
\textbf{Homomorphic encryption} (HE~\cite{chillotti_2020_tfhe, gentry_2009_fully, cheon_2017_homomorphic, fan_2012_somewhat}): HE is a set of public-key encryption schemes that allow arithmetic operations on the ciphertext without the need for decryption.
Assuming messages $m_1$ and $m_2$, a public-key $pk$, a secret key $sk$, an HE encryption scheme $\mathbb{E}$, and an HE decryption scheme $\mathbb{D}$, a homomorphic evaluation $Eval$ satisfies the following upon some function of $f$:
$\mathbb{D}(sk, Eval(pk,\mathbb{E}(pk,m_1),\mathbb{E}(pk,m_2),f))\!=\!f(m_1,m_2)$.
Due to the HE nature that errors accumulate over computations, only a limited number of HE operations ($Eval$) can be executed upon an encrypted message.
A bootstrapping operation can reset the accumulated errors, but only at the cost of high computational complexity.
A leveled-HE refers to a set of HE schemes that do not support bootstrapping operation incurring lower cost but providing limited depth, while fully-HE refers to those that support the bootstrapping.

\begin{table}[tb!]
    \centering
    \caption{Latency and communication costs of BT processing polynomial activation and GC processing ReLU. Each 
    is measured in the Delphi protocol for ResNet32, amortized to a single operation.}
    \vspace{0.1in}
    \label{tbl:2_GC_BT_cost}
    \resizebox{0.48\textwidth}{!}{
    \begin{tabular}{l|l|l|l|l} 
    \Xhline{3\arrayrulewidth}
    \multirow{2}{*}{\textbf{Activation function}} & \multicolumn{2}{l|}{\textbf{Time} (µs)} & \multicolumn{2}{l}{\textbf{Comm.} (KB)}  \\ 
    \cline{2-5}
                                         & Offline & Online               & Offline & Online                        \\ 
    \Xhline{2\arrayrulewidth}
    Polynomial activation                & 2.80    & 1.20                 & 0.192   & 0.036                         \\ 
    \hline
    ReLU                                 & 60.60   & 20.22                & 19.088  & 1.184                         \\
    \Xhline{3\arrayrulewidth}
    \end{tabular}

    }
    \vspace{-0.2in}
\end{table}

GC supports arbitrary functions such as ReLU, but only at the cost of high communication/storage for the garbled circuits and exchanged labels.
BT supports multiplications with relatively lower-cost than GC but can only support polynomial activation functions, which leads to lower model accuracy.
Table~\ref{tbl:2_GC_BT_cost} summarizes the amortized latency and communication costs of BT and GC, while each processing polynomial activation and ReLU, respectively.
HE is different in that it does not require multiple communication rounds or additional storage costs; however, it is limited in the depth of the computation and the supportable operations such as compare, which is necessary for ReLU.

\begin{figure*}[tb!]
    \begin{center}
    \includegraphics[width=1.85\columnwidth]{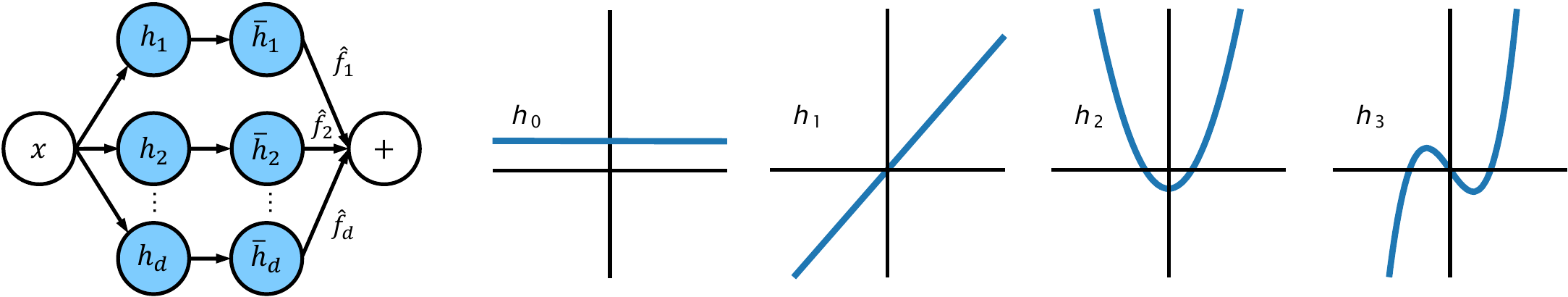}
    \end{center}
    \vspace{-0.12in}
    \caption{HerPN block. For one input feature, several Hermite basis polynomials of input are evaluated. The computed Hermite polynomials experience separate basis-wise normalization and weighted summation using Hermite coefficients (left). The first four Hermite basis polynomials are depicted in the right.}
    \label{fig:3_herpn}
    \vspace{-0.11in}
\end{figure*}

\subsection{Delphi overview}
\label{sec:2_delphi_overview}

Delphi~\cite{lehmkuhl_2020_delphi} is a representative hybrid PI protocol based on the MPC primitives and HE.
Delphi exploits LHE for the linear layers of an ML model and either GC or BT for the activation functions of the model.
The inference under the Delphi protocol consists of offline and online phases (see Figure~\ref{fig:2_delphi_overview} in Appendix~\ref{sec:appendix_a}).
The offline phase is independent of the client’s actual input data and thus can be pre-processed.
The online phase begins when the client sends their private data to the server and ends when the client learns the final inference results. 


\noindent
\textbf{Linear layer}: During the offline phase, random matrices ($r_i$ and $s_i$ for $i^{th}$ layer at the client and server) are generated. Using the encypted $\mathbb{E}(r_i)$ from the client, the server computes $\mathbb{E}(W_i r_i\!-\!s_i)$ using the HE and returns it. The client decrypts and acquires $W_i r_i\!-\!s_i$. When the online phase starts, the server computes the plaintext operation using $x_i\!-\!r_i$ (SS of input $x_i$) to obtain $W_i(x_i\!-\!r_i)\!+\!s_i$, which results in the SS of $W_i x_i$ between the client and the server.

\noindent
\textbf{Activation layer}: Cryptographic primitives used for the activation layer depend on the type of the activation function; GC for ReLU or BT for polynomial activation, quadratic approximation in particular. In the case of i) \emph{ReLU and GC}, the server acts as a garbler and the client acts as an evaluator. During the offline phase, the garbled circuit $\hat{C}$ for ReLU, labels $L(r_{i+1})$, and $L(W_i r_i\!-\!s_i)$ are sent from the server to the client. At the online phase, $L(W_i (x_i\!-\!r_i)\!+\!s_i)$ is sent from the server to the client, where $\hat{C}$ is evaluated and generates $x_{i+1}\!-\!r_{i+1}$, sending it back to the server. 
In contrast, for ii) \emph{quadratic approximation and BT}, the server and client simply secret-share Beaver’s triples during the offline phase. At the online phase, they jointly compute and gain the secret share of $x_{i+1}$ consuming a set of triples; $\langle x_{i+1}\rangle_1$ and $\langle x_{i+1}\rangle_2$ for the client and server. The client then sends $\langle x_{i+1}\rangle_1\!-\!r_{i+1}$ to the server, allowing the server to gain $x_{i+1}\!-\!r_{i+1}$ for the next linear layer in the online phase.

Delphi employs NAS~\cite{zoph_2017_neural} to search for an optimal ML model that replaces some of its GC-processed ReLUs with BT-processed quadratic approximations, considering the tradeoff between the model accuracy and costs, including the online serving latency.
The online latency is dominated by the activation layers in particular (see Figure~\ref{fig:1_time_breakdown}).
While the linear layer only requires fast plaintext computation, the activation layer demands high communication costs between the client and the server, which are especially higher in the case of GC compared to BT.
However, one cannot na\"ively replace ReLUs (processed via GC) with polynomial activations (processed via BT) to reduce the online latency because the accuracy of an ML model easily deteriorates when the ReLU functions are replaced by simple quadratic approximation functions. 
Delphi attempts to find a sweet spot amongst this tradeoff exploiting NAS. 

Although promising, the latency of the obtained ML models still shows large room for improvements (taking 11.3 secs for inferring ResNet18 on CIFAR-10 in our setup) compared to 0.1354 msec \cite{coleman_2017_dawnbench} in a non-PI context.
Multiple works including \cite{lou_2020_safenet, cho_2021_sphynx, ghodsi_2020_cryptonas, jha_2021_deepreduce} further optimized the training method, NAS structure, or even the polynomial activation function options, but these works have still achieved limited success in the latency or accuracy.

\subsection{Activation Function Approximation}
\label{sec:2_poly_functions}

Various prior arts have attempted to provide effective polynomial activation functions~\cite{garimella_2021_sisyphus,ishiyama_2020_highly,obla_2020_effective,hesamifard_2019_deep,thaine_2019_efficient,lou_2021_hemet,lee_2021_precise, gilad_2016_cryptonets}.
However, they either suffer from the inferior \emph{accuracy} of the model or the high online service \emph{latency} in the PI context. 
Especially when using low degree polynomials (2 or 3), they can only train shallow neural networks consisting of fewer than ten layers. \cite{garimella_2021_sisyphus} proposed a QuaIL method that trains a polynomial activation-based model by optimizing the mean squared error (MSE) of an intermediate result from a pretrained ReLU-based model and report training deep networks using degree 2 polynomials.
However, the reported accuracy is noticeably inferior to that of neural networks using ReLU.
Contrarily, \cite{lee_2021_precise} proposed a high degree (e.g., 29) minimax polynomial approximation of ReLU and max-pooling with minimal accuracy drop for ImageNet. 
However, high degree polynomials are costly to evaluate \cite{lee_2021_precise} on the PI framework with higher communication and computational costs.

Circa~\cite{ghodsi_2021_circa} refactored the ReLU function into a piece-wise linear function and a sign function, utilizing BT for the former and GC for the latter.
In particular, Circa further approximated the sign function with stochastic sign and trimming of the input, through which it minimized the size and thus cost of the boolean circuit to be evaluated via GC.
Although Circa demonstrated superior (lower) latency over prior studies with comparable serving accuracy, there still exists large room for improvements.

\section{Hermite polynomial with basis-wise normalization}
\label{sec:3_contribution}



\subsection{Orthogonal basis and Hermite polynomials}
\label{sec:3_orthogonal_hermite}

Our work utilizes Hermite expansion, a Fourier transform using Hermite polynomials as eigen-functions.
Before we go deep into the construction of our activation function, we briefly describe the basic properties of orthogonal bases and Hermite polynomials.
For a real interval $[a,b]$, let $L^2([a,b],w(x))$ be the space of square-integrable functions with respect to a weight function $w(x)$.
Square-integrable function is $f:\mathbb{R} \rightarrow \C$ such that
\begin{equation}
     \int_{a}^{b} {|f(x)|^2w(x) \,dx}<\infty
\end{equation}
Then $L^2([a,b],w(x))$ is a Hilbert space where the inner product is defined.
The inner product of two functions $f,g \in L^2([a,b],w(x))$ is defined as follows:

\begin{equation}
    \langle f,g \rangle =  \int_{a}^{b} {f(x)\overline{g(x)}w(x)
    \,dx} 
\end{equation}

\noindent A finite set of polynomials $\{p_1(x), p_2(x), \dots, p_n(x)\}$ forms an orthogonal basis if the set spans $L^2([a,b],w(x))$ and 
$\langle p_i, p_j\rangle=0$ if $i \neq j $.

Hermite polynomials are a type of orthogonal polynomials that arise in probability and physics.
While there are other sets of polynomials that form an orthogonal basis such as Chebyshev or Laguerre, Hermite polynomials especially feature i) Gaussian weight functions defined in $\mathbb{R}$ and ii) orderedness, as the low order parts of the Hermite expansion hold most of the information.
Mathematically, probabilist’s Hermite polynomials  are given by: 

\begin{equation}
    H_n(x)=(-1)^n e^{\frac{x^2}{2}}\frac{d^n}{dx^n}e^{\frac{-x^2}{2}}
    \label{eq:3_hermite_polynomial}
\end{equation}

\noindent Then, the normalized Hermite polynomials $h_n(x)=\frac{1}{\sqrt{n!}} H_n(x)$ form an orthonormal basis in the $L^2(\mathbb{R},e^{-x^2/2})$ in the sense that the series of the polynomial $\{h_i\}_{i=0}^{\infty}$ are orthonormal: $\langle h_i,h_j \rangle = \delta_{i,j}$. Here $\delta_{i,j}=1 $ if $i=j$ and otherwise $\delta_{i,j}=0$

Finally, given a function $f \in L^2(\mathbb{R},e^{-x^2/2})$, we have the Hermite expansion defined as the following:

\begin{equation}
    f(x) = \sum_{i=0}^{\infty}\hat{f_i}h_i(x), \quad \hat{f_i} = \langle f,h_i \rangle
    \label{eq:4_hermite_coefficient}
\end{equation}

\noindent ${\hat{f_i}}$ is the $i$-th Hermite coefficient of $f$, defined as the inner product of the function $f$ and the Hermite polynomial $h_i$. 

While some prior works exploited Hermite polynomials in the neural networks, none had sought the use in the context of PI.
\cite{panigrahi_2020_effect,agarwal_2021_deep,ge_2017_learning} exploited the Hermite expansion of activation functions in a theoretical domain.
\cite{lokhande_2020_generating} demonstrated that the Hermite expansion of ReLU with a soft-sign function shows fast training loss convergence for a pseudo labeling task.

\begin{figure}[tb!]
    \begin{center}
    \includegraphics[width=1\columnwidth]{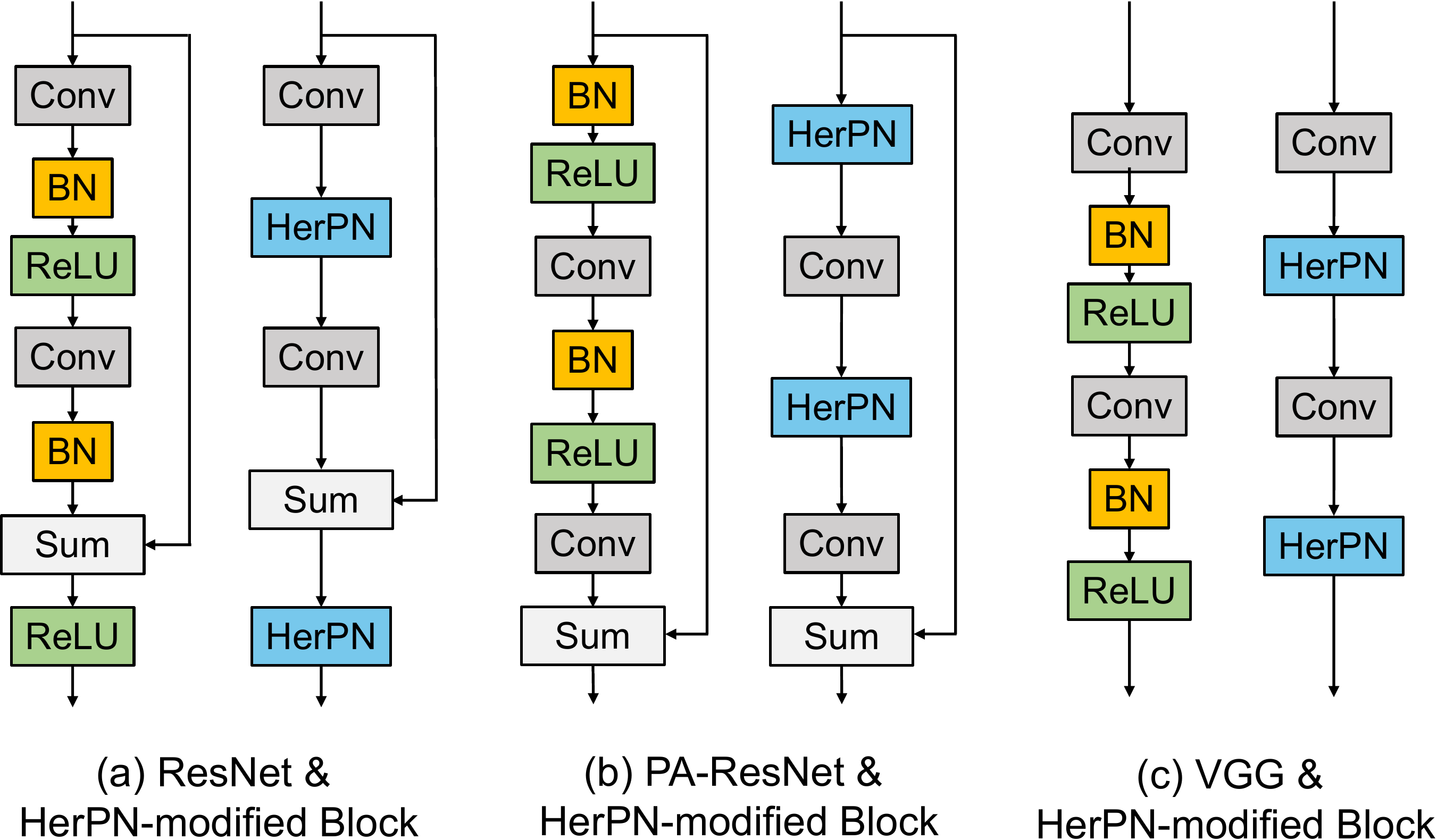}
    \end{center}
    \vspace{-0.15in}
    \caption{ResNet, PA-ResNet, and VGG ML architectures that are modified using HerPN blocks.}
    \label{fig:3_replaced_models}
    \vspace{-0.1in}
\end{figure}

\subsection{The HerPN method}
\label{sec:3_herpn_method}

We propose the Hermite expansion of ReLU with basis-wise normalization (\emph{HerPN}) block (see Figure~\ref{fig:3_herpn}) as a substitute for the ReLU and the normalization functions in neural networks.
The $n$-th Hermite coefficient of the ReLU, $\hat{f_n}$ is: 
\begin{equation*}
    \hat{f_n}= \begin{cases}
    \frac{1}{\sqrt{2 \pi}} &\text{$n = 0$},\\
    \frac{1}{2} &\text{$n = 1$},\\
    0 &\text{$n \geq 2$ and odd},  \\
    \frac{((n-3))!!)^2}{\sqrt{2 \pi n!}} &\text{$n \geq 2$ and even}
    \end{cases}
\end{equation*}
\vspace{-0.1in}


\begin{table*}[tb!]
    \centering
    \caption{\name online latency and total communication cost comparison with the baseline neural networks in Delphi for CIFAR-100 (C100) and TinyImageNet (Tiny). As the networks are equally structured while processing CIFAR-10 and CIFAR-100 except last fully connected layer, we only demonstrate CIFAR-100 between the two. Full accuracy results in Appendix~\ref{sec:appendix_c}.
    }
    \vskip 0.15in
    \label{tbl:4_delphi_comparison}
    \resizebox{0.8\textwidth}{!}{
    \begin{tabular}{l |c|c |r |c|c|c|c|c S[table-format=4.1] S[retain-zero-exponent=true]}
        \Xhline{3\arrayrulewidth}
        \multirow{2}{*}{\textbf{Network}} & \multicolumn{2}{c|}{\textbf{Accuracy}(\%)} & \multicolumn{3}{c|}{\textbf{Online Latency} (ms)} & \multicolumn{3}{c}{\textbf{Communication} (MB)}  \\ 
        \cline{2-9}
                    & Baseline & \name & Delphi & \name & Improv      & Delphi  & \name & Improv           \\ 
        \hline
        VGG16-C100       & 73.45 & 71.99 & 6140.1   & 209.3 & 29.3$\times$            & 5782.3 & 240.9 & 24.0$\times$                 \\
        ResNet18-C100    & 77.93 & 77.40 & 12281.8  & 291.6 & 42.1$\times$            & 11699.5 & 533.8 & 21.9$\times$                 \\
        ResNet32-C100    & 71.66 & 63.98 & 6604.7   & 386.0 & 17.1$\times$            & 6403.8 & 328.4 & 19.5$\times$                 \\
        PA-ResNet18-C100 & 76.95 & 76.31 & 12205.3  & 405.1 & 30.1$\times$            & 11533.4 & 532.0 & 21.7$\times$                 \\
        PA-ResNet32-C100 & 70.21 & 67.85 & 7443.9 & 373.1 & 23.4$\times$            & 7349.1 & 371.4 & 19.8$\times$                 \\
        \hline
        VGG16-Tiny       & 60.80 & 58.84 & 24232.5   & 505.7 & 47.9$\times$            & 22960.4 & 793.4 & 28.9$\times$                 \\
        ResNet18-Tiny    & 63.72 & 63.35 & 48887.4  & 814.8 & 60.0$\times$            & 46685.7 & 2658.2 & 17.6$\times$                 \\
        ResNet32-Tiny    & 55.06 & 43.04 & 26320.9   & 621.7 & 42.3$\times$            & 25408.8 & 1446.4 & 17.6$\times$                 \\
        PA-ResNet18-Tiny & 61.95 & 61.50 & 48587.5  & 791.9 & 61.4$\times$            & 46021.4 & 2015.8 & 22.8$\times$                 \\
        PA-ResNet32-Tiny & 55.58 & 53.09 & 30535.8 & 647.1 & 47.2$\times$            & 29189.9 & 1275.9 & 22.9$\times$                 \\
        \Xhline{3\arrayrulewidth}
    \end{tabular}
    }
    \vspace{-0.1in}
\end{table*}

Based on Formula~\ref{eq:4_hermite_coefficient}, we use three bases with the highest degree of 2 ($h_0$, $h_1$, and $h_2$ in Figure~\ref{fig:3_herpn}) for the HerPN block.
Using four bases shows similar or slightly lower accuracy compared to the case of using three. 

We also employ basis-wise normalization concatenated to each basis, instead of a standard pre/post-activation normalization.
Basis-wise normalization computes the mean and variance of each Hermite polynomial over the mini-batch of training data.
We use Hermite coefficients as fixed weight and place scale and shift parameters after accumulating the Hermite polynomials.
To sum up, HerPN can be computed on input $x$ as follows:

\vspace{-0.1in}
\begin{equation}
    f(x) = \gamma \sum_{i=0}^{d}\hat{f_i}\frac{h_i(x)-\mu}{\sqrt{\sigma^2+\epsilon}}+\beta
\end{equation}
\vspace{-0.1in}

Basis-wise normalization is critical in attaining the desired accuracy of the model for the following reasons.
First, without a proper normalization technique, the output of the polynomial activation during the forward or backward propagation can exponentially increase.
Thus, the ranges of a mini-batch in each layer can change drastically. Such a characteristic exacerbates with the high-degree polynomials or deep neural networks, potentially leading to the exploding or vanishing gradient problem~\cite{garimella_2021_sisyphus, gilad_2016_cryptonets}.

Second, without the basis-wise normalization, the post-activation value may be dominated by a single highest or lowest order Hermite basis due to the differences in the size of values from the polynomial degrees.
The standard post- or pre-activation normalization does not sufficiently alleviate these problems as they overlook the scale difference between the basis expressions.
Basis-wise normalization is advantageous as it keeps each basis zero-centered and consistently keeps the range of the intermediate values throughout the layers.

Using the HerPN block, we can directly replace the batch-normalization (BN) and ReLU blocks in the popular ML architectures such as pre-activation ResNet (PA-ResNet~\cite{he_2016_identity}) and VGGNet~\cite{simonyan_2014_very} (Figure~\ref{fig:3_replaced_models}(b), (c)).
We cannot directly apply HerPN to ResNet~\cite{he_2016_deep} because a ReLU function exists after the skip-connection.
We thus modify the ResNet model as shown in Figure~\ref{fig:3_replaced_models}(a).
Although we only demonstrate the HerPN block based on the Hermite expansion of the ReLU function, our method can also be applied to other activation functions, such as ELU~\cite{clevert_2015_fast}, SELU~\cite{klambauer_2017_self}, Swish~\cite{ramachandran_2017_searching}, GeLU~\cite{hendrycks_2016_gaussian}, and Mish~\cite{misra_2019_mish}.

\section{Evaluation}
\label{sec:4_evaluation}

\subsection{Experimental setup}
\label{sec:4_experimental_setup}

\noindent
\textbf{Benchmarks and datasets}: We tested the private inference of our trained neural networks on CIFAR-10/100~\cite{krizhevsky_2009_cifar} and TinyImageNet~\cite{yao_2015_tiny} datasets.
CIFAR-10/100 (C10/100) datasets consist of 32$\times$32 50,000 training and 10,000 test images. 
The only difference between the two datasets is the number of classes (10 for C10 and 100 for C100).
TinyImageNet dataset has higher resolution of 64$\times$64 and consists of 100,000 training images and 10,000 test images with 200 output classes.

We performed experiments on VGG16, ResNet18/32, and  PA-ResNet18/32.
We applied our HerPN method to these networks and tailored them to the evaluated datasets by adjusting the pooling or fully-connected layers prior to the final softmax as \cite{garimella_2021_sisyphus}.
We first compared the performance of our HerPN-based ML models with the baseline Delphi (Section~\ref{sec:4_compare_delphi}).
We also compared our HerPN-based models with the prior polynomial activation works of Lee et.al.~\cite{lee_2021_precise} and QuaIL~\cite{garimella_2021_sisyphus} using CIFAR-10 (Section~\ref{sec:4_compare_prior_poly}).

\noindent
\textbf{System setup}: We used the AWS system to evaluate the effectiveness of our HerPN method on Delphi, obtaining the latency and communication/storage usage measurements.
The hardware configurations of the client and the server are both AWS c5.4xlarge instances for CIFAR-10/100, which consist of Intel Xeon 8000 series CPU at 3.0 GHz with 32 GB of RAM. We further empolyed AWS c5.9xlarge instances equipped with the same CPU, each having a 72GB of RAM for TinyImageNet.
The client and server instances were both located in the ap-northeast-2 regions.
The communication link between the client and server was in the LAN setting.
We used the open-source Delphi version using the SEAL library~\cite{sealcrypto} for HE and the fancy garbling~\cite{fancygarbling} for GC. 
We report the average latency and communication cost results of 10 different experiments, which shows less than 5\% fluctuation in their results.

\subsection{Effectiveness of the HerPN-based models on the Delphi protocol}
\label{sec:4_compare_delphi}

ML models with HerPN blocks demonstrate superior online latency (ranging from 17.1$\times$ to 61.4$\times$) and communication costs (ranging from 17.6$\times$ to 28.9$\times$) for all the evaluated ML models compared to the baseline Delphi (see Table~\ref{tbl:4_delphi_comparison}).
\name shows large runtime improvement over the prior PI protocols.  
SAFENet~\cite{lou_2020_safenet}, which further employed an additional channel-wise NAS, demonstrated 2.5$\times$/2.46$\times$ reduction in online latency for ResNet32/C100 and VGG16/C10 over the Delphi protocol. 
Circa, a state-of-the-art protocol that utilized stochastic ReLU, demonstrated 2.6$\times$/2.6$\times$ runtime improvement for the same ResNet32/C100 and VGG16/C10,
while those of \name are 17.1$\times$ and 29.3$\times$ over Delphi.

This large reduction of costs is primarily attributed to the fact that HerPN block can replace all ReLUs in the original Delphi, allowing the full usage of the BT instead using the costly GC primitive.
Although we only evaluated the HerPN-modified neural networks on Delphi, \name are applicable to other PI protocols~\cite{gilad_2016_cryptonets, juvekar_2018_gazelle} that are limited by activation functions.

\subsection{Comparison with the prior activation functions}
\label{sec:4_compare_prior_poly}


ML models with the HerPN blocks show accuracies comparable to those of the baseline neural networks with ReLU. Table~\ref{tbl:4_delphi_comparison} reports the accuracy of the trained baseline and HerPN-based models (Appendix~\ref{sec:appendix_c} shows the full result).
In the case of VGG16, ResNet18 and PA-ResNet18/32, our work shows comparable accuracy.
HerPN-based ResNet32 shows a large accuracy drop on CIFAR-100 and TinyImageNet because HerPN cannot natively support the original ResNet with the skip connection. 
However, such drawbacks can be nullified considering that the accuracy of the HerPN-based PA-ResNet32 is higher than that of the original ResNet32.

\begin{table}[tb!]
    \centering
    \caption{ML model accuracy and degree comparison of \name with prior polynomial activation works on CIFAR-10.}
    \vspace{0.1in}
    \label{tbl:4_prior_poly}
    \resizebox{0.48\textwidth}{!}{
    \begin{tabular}{l|l|l|l}
        \Xhline{3\arrayrulewidth}
        \textbf{Method}                    & \textbf{Neural Network} & \textbf{Accuracy} & \textbf{Degree} \\ 
        \hline
        \multirow{3}{*}{\name}              & VGG16          & 92.38\%         & 2       \\
                                           & ResNet18       & 94.96\%         & 2       \\
                                           & ResNet32       & 93.83\%         & 2       \\ 
        \hline
        \multirow{2}{*}{\cite{lee_2021_precise}} & VGG16          & 91.87\%         & 29      \\
                                           & ResNet32       & 89.23\%         & 29      \\ 
        \hline
        \multirow{3}{*}{QuaIL}          & VGG16          & 82.25\%         & 2       \\
                                           & ResNet18       & 83.61\%         & 2       \\
                                           & ResNet32       & 71.81\%         & 2       \\
        \Xhline{3\arrayrulewidth}
    \end{tabular}
    }
    \vspace{-0.2in}
\end{table}

Table~\ref{tbl:4_prior_poly} compares the modified VGGNet and ResNet implemented by the prior polynomial activation works and HerPN on CIFAR-10.
The accuracies of the HerPN-based models are consistently higher than QuaIL, which uses the same degree of quadratic polynomials.
Compared to \cite{lee_2021_precise}, as the degree of polynomials is far smaller, our work has significantly less computational cost on activation functions.
Yet, \name shows accuracy consistently higher than that of \cite{lee_2021_precise} on CIFAR-10.

\begin{figure*}[tb!]
    \begin{center}
    \includegraphics[width=1.75\columnwidth]{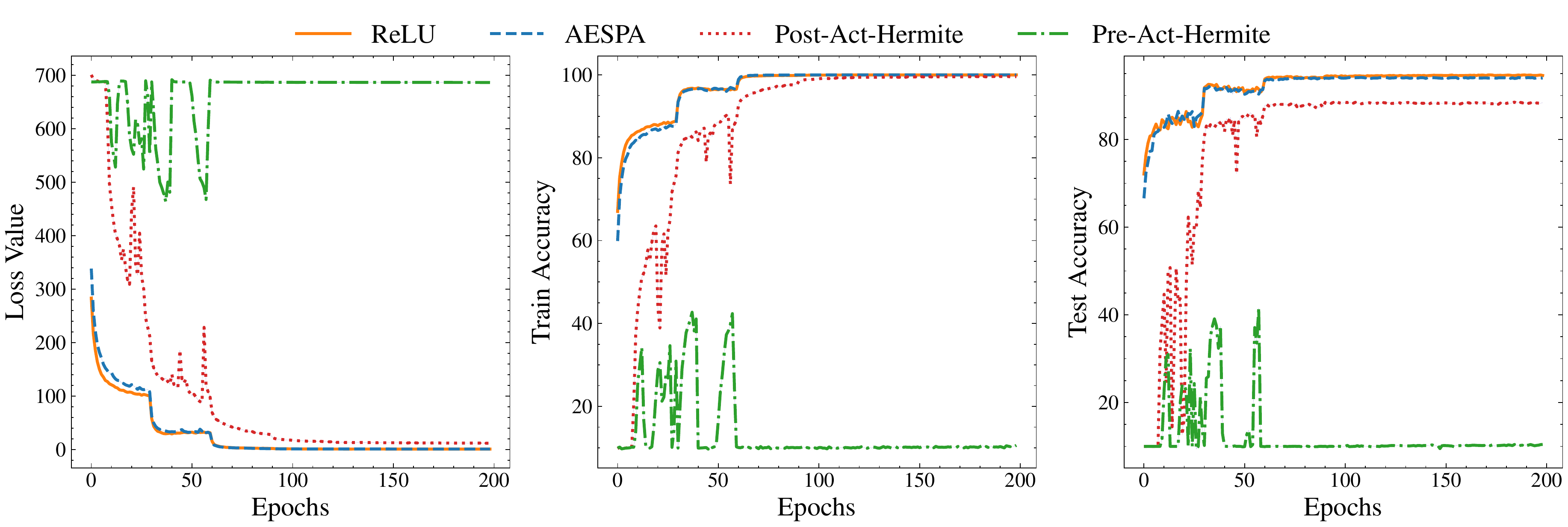}
    \end{center}
    \vspace{-0.1in}
    \caption{Comparing ML models on CIFAR-10 using different normalization techniques: i) baseline original ResNet32 with ReLU, ii) HerPN-based model, and ResNet32 with Hermite polynomial activation but with iii) post-act normalization, and iv) pre-act normalization.}
    \label{fig:4_norm_ablation}
    \vspace{-0.1in}
\end{figure*}

\subsection{Effectiveness of basis-wise normalization}
\label{sec:4_ablation_study}

Through an ablation study regarding the normalization technique for the HerPN block, we identified that our basis-wise normalization shows a clear advantage in terms of model accuracy over other options.
Besides the baseline ResNet32 with ReLU activation and the HerPN-based ResNet32, we trained two additional ResNet32 models that exploit the Hermite expansion of the ReLU as the activation function, but adopt standard i) \emph{pre-activation} normalization and ii) \emph{post-activation} normalization, instead of our basis-wise normalization. 
Figure~\ref{fig:4_norm_ablation} illustrates the loss value, train accuracy, and test accuracy of the four trained models.
In the case of the pre-activation normalization, the loss does not converge to zero and fluctuates during the training.
While the post-activation normalization does achieve high train accuracy, the HerPN-based case shows much higher \emph{test} accuracy with a minimum accuracy drop compared to the baseline case with ReLU.

\section{Related work}
\label{sec:5_related_work}

\subsection{Reducing the serving cost of activation functions for PI protocols}
\label{sec:5_relu_reduction}

Gazelle~\cite{juvekar_2018_gazelle} solely relies on GC and additive masking in calculating the non-linear functions.
Delphi reduces the number of ReLUs relying on GC via NAS.
Given the pre-trained neural network, Delphi replaces some ReLUs with quadratic approximation and retrains the neural network, which allows them to utilize a less costly BT primitive.
The actual replacement process exploits a planner conducting population-based training (PBT~\cite{jaderberg_2017_population}) that maximizes the number of ReLU functions to be replaced with quadratic approximation and minimizes the accuracy drop.
SAFENet~\cite{lou_2020_safenet} extends the replacement policy of Delphi.
SAFENet introduces a more fine-grained channel-wise replacement that picks activation between zero pruning and polynomials of degree 2 or 3 for each channel.

The key weakness of these replacement-based approaches is that even with the NAS-based planner, polynomial approximation results in a severe accuracy drop after a certain approximation ratio.
CIFAR-100 accuracy of ResNet32 using Delphi’s planner provides less than a 2\% accuracy drop within 26 ReLU function approximations out of 31 ReLU functions.
However, the accuracy of a network applying approximation to 27 or more ReLU functions drops quickly compared to the network without approximation.
Considering that even a small number of ReLU activation layers with GC incur the majority of the overall cost, the fact that not all activation layers can be approximated is a critical bottleneck.

Another approach is to design neural architectures from scratch optimized for the ReLU counts.
\cite{jha_2021_deepreduce} searches for neural architectures that can efficiently minimize the number of invoking ReLUs, rather than total FLOPs in typical NAS~\cite{zoph_2017_neural}.
However, despite using these optimizations, the remaining ReLUs account for most of the online latency.

\subsection{Polynomial activation function works}
\label{sec:5_poly_function}

Since CryptoNet~\cite{gilad_2016_cryptonets}, there have been challenges to search for polynomial activations for PI. \cite{ishiyama_2020_highly,obla_2020_effective,hesamifard_2019_deep,thaine_2019_efficient,lou_2021_hemet,lee_2021_precise} exploit classical numerical approximations to implement polynomials that can act as the original activation function.
The resulting polynomial activations are based on Taylor, Chebyshev, or MinMax approximations.
These works have demonstrated superior accuracy to the square function.
However, only two of them demonstrated the effectiveness of their work on deep neural networks.

First, \cite{lee_2021_precise} approximates a ReLU function with a high degree polynomial that successfully preserves the original model accuracy and proved its trainability on deep neural networks such as ResNet50~\cite{he_2016_identity} and ImageNet\cite{russakovsky_2015_imagenet}.
However, the high degree of the polynomial leads to longer latency on the hybrid PI protocol framework even with the BT because it necessitates a greater number of MPC primitive usages.

\cite{garimella_2021_sisyphus} uses quadratic polynomial approximations and introduces QuaIL.
QuaIL is an alternative training method that minimizes the MSE loss between the intermediate representations of a pre-trained ReLU-based neural network and a target quadratic approximation-based neural network.
With QuaIL, one can avoid exploding-gradient problems as they train a single layer at a moment while the others are frozen.
However, QuaIL failed to achieve competitive accuracy because even small differences in the intermediate representation can cause divergence in subsequent layers.

Each study showed the potential of polynomials as activation functions. 
However, there has been no prior work demonstrating accuracy competitive to the baseline ReLU using low-degree polynomials.

\section{Conclusion}
\label{sec:6_conclusion}
 
In this paper, we have proposed an accuracy preserving low-degree polynomial activation function, \name.
\name leverages the Hermite polynomial concatenated with basis-wise normalization, which retains the high accuracy of the standard ML models using ReLU.
Our proposal is especially effective in the PI context because i) the polynomial activation enables less expensive BT primitive instead of GC, ii) the low degree of our work reduces the number of MPC primitives required for serving, and iii) \name demonstrates a high ML model accuracy even without the need of complex NAS or training methods.
We also showcase the effectiveness of our work with the open-source PI protocol, Delphi, on a real machine and network for VGGNet, ResNet, and pre-activation ResNet; the online latency and the communication between the client and the server are reduced by up to 61.4$\times$ and 28.9$\times$, respectively.

\balance
\bibliographystyle{IEEEtran}
\bibliography{reference}

\newpage
\appendices
\onecolumn
\section{An illustration of the Delphi protocol}
\label{sec:appendix_a}

\begin{figure*}[h]
    \begin{center}
    \includegraphics[width=0.9\columnwidth]{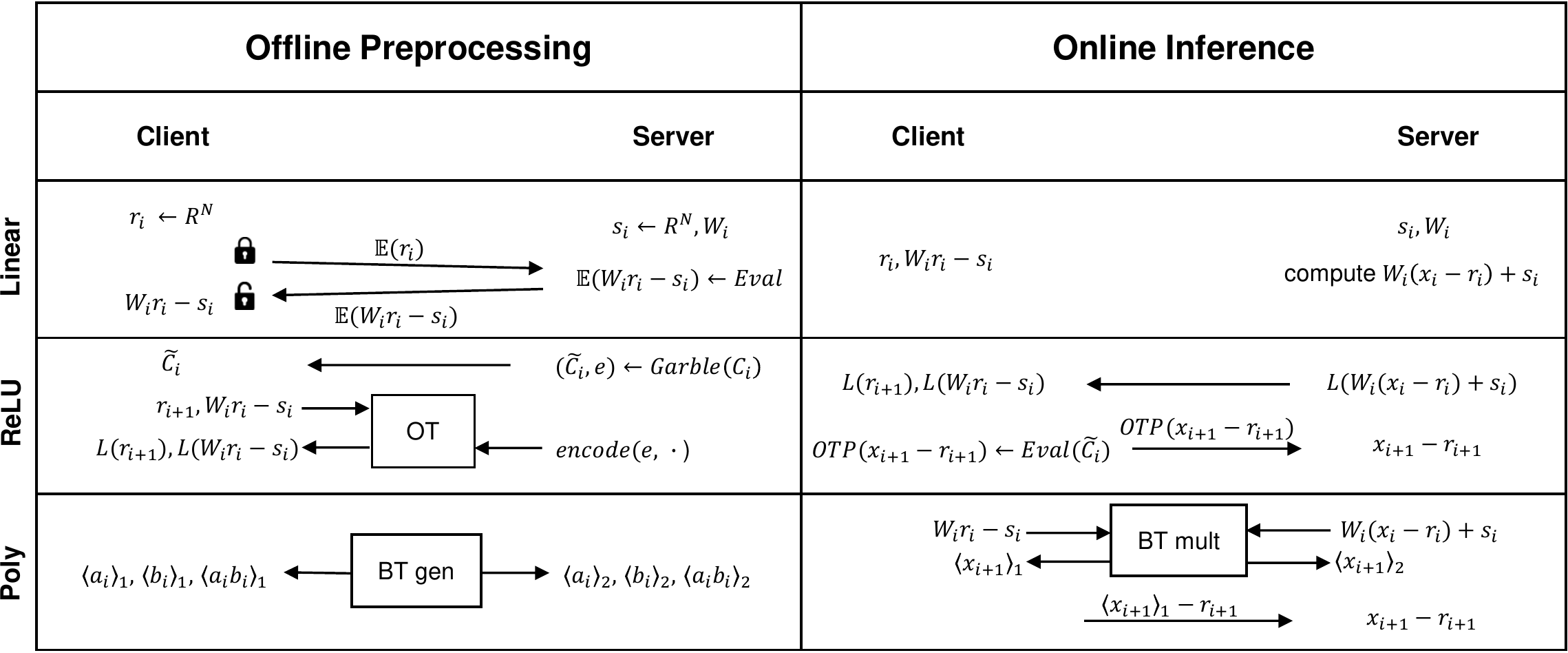}
    \end{center}
    \vspace{-0.2in}
    \caption{An illustration of the Delphi protocol. Delphi uses LHE for convolution and FC in linear layers, 
    and GC for ReLU or BT for polynomial activation in non-linear layers.}
    \label{fig:2_delphi_overview}
    \vspace{-0.1in}
\end{figure*}

\section{Parameters for cryptographic primitives}
\label{sec:appendix_b}
\noindent
\textbf{Choice of prime for prime field}: In Delphi, the underlying MPC cryptographic primitives have two additional parameters: i) \emph{prime number for prime field} and ii) \emph{data representation}. All the operations in the cryptographic primitives are defined on this data representation and prime finite field. This parameter setting ensures no overflow and underflow to occur in a single layer evaluation. Throughout this paper, we have used the latest version of Delphi 's configuration\cite{Delphi}, which uses 41-bit prime of 2061584302081 for prime finite fields and 11bit fixed-point data representation.

\section{Training results}
\label{sec:appendix_c}

\noindent
Models presented in the paper are trained normally using PyTorch. More precisely, we trained all networks for 200 epochs using SGD with cosine annealing scheduler, 0.1 initial learning rate, 100 batch size, 0.0005 weight decay, 0.9 momentum. Table \ref{tbl:4_baseline_accuracy_comparison} summarizes training accuracy of AESPA and baseline ReLU networks in CIFAR-10/100 and TinyImageNet. We adopt standard data augmentations: random crops, random flips, random rotation and normalization. Soft labeling is used to obtain higher accuracy.

\begin{table}[h!]
    \centering
    \caption{Accuracy result of \name on CIFAR-10 (C10), CIFAR-100 (C100) and TinyImageNet (Tiny), compared to the standard ML models with ReLU and batch-normalization.}
    \vspace{0.1in}
    \label{tbl:4_baseline_accuracy_comparison}
    \resizebox{0.44\textwidth}{!}{
    \begin{tabular}{l|l|l|l} 
        \Xhline{3\arrayrulewidth}
        \textbf{Dataset}      & \textbf{Neural Network} & \textbf{ReLU Acc} & \textbf{\name Acc} \\ 
        \hline
        \multirow{5}{*}{C10}  & VGG16        & 93.95\%             & 92.38\%               \\
                              & ResNet18     & 95.57\%             & 94.96\%               \\
                              & ResNet32     & 93.85\%             & 93.83\%               \\
                              & PA-ResNet18 & 94.75\%             & 94.59\%               \\
                              & PA-ResNet32 & 93.21\%             & 91.79\%               \\ 
        \hline
        \multirow{5}{*}{C100} & VGG16        & 73.45\%             & 71.99\%               \\
                              & ResNet18     & 77.93\%             & 77.40\%               \\
                              & ResNet32     & 71.66\%             & 63.98\%               \\
                              & PA-ResNet18 & 76.95\%             & 76.31\%               \\
                              & PA-ResNet32 & 70.21\%             & 67.85\%               \\
        \hline
        \multirow{5}{*}{Tiny} & VGG16        & 60.80\%             & 58.84\%               \\
                              & ResNet18     & 63.72\%             & 63.35\%               \\
                              & ResNet32     & 55.06\%              & 43.04\%               \\
                              & PA-ResNet18 & 61.95\%               & 61.50\%               \\
                              & PA-ResNet32 & 55.58\%               & 53.09\%               \\
        \Xhline{3\arrayrulewidth}
    \end{tabular}
    }
\end{table}

\section{Comparison with Prior Work}
\label{sec:appendix_d}

\noindent
Table \ref{tbl:5_vgg16_baseline_speedup_comparison} and \ref{tbl:6_resnet32_baseline_speedup_comparison} shows the accuracy and speedup result for VGG16 on CIFAR-10 and ResNet32 on CIFAR-100. While \name shows almost an order of magnitude speedup over all prior works, \name's accuracy result is not the highest. However, considering the larger network such as PA-ResNet18 in our experiments, \name outperforms all prior works in terms of accuracy and speedup. 

\begin{table}[h!]
    \centering
    \caption{The VGG16 result on CIFAR-10.}
    \vspace{0.1in}
    \label{tbl:5_vgg16_baseline_speedup_comparison}
    \resizebox{0.4\textwidth}{!}{
    \begin{tabular}{l|l|l} 
        \Xhline{3\arrayrulewidth}
        \textbf{VGG16} & \textbf{Accuracy} & \textbf{speedup} \\ 
        \hline
        Delphi  & 88.1\%             & 1.2$\times$                \\
        SAFENet & 88.9\%             & 2.5$\times$                \\
        Circa   & 93.8\%             & 2.6$\times$                \\
        AESPA   & 92.4\%             & 24.0$\times$                \\ 
        \Xhline{3\arrayrulewidth}
    \end{tabular}
    }
\end{table}

\begin{table}[h!]
    \centering
    \caption{The ResNet32 result on CIFAR-100.}
    \vspace{0.1in}
    \label{tbl:6_resnet32_baseline_speedup_comparison}
    \resizebox{0.4\textwidth}{!}{
    \begin{tabular}{l|l|l} 
        \Xhline{3\arrayrulewidth}
        \textbf{ResNet32} & \textbf{Accuracy} & \textbf{speedup} \\ 
        \hline
        Delphi  & 67.3\%             & 1.3$\times$                \\
        SAFENet & 67.5\%             & 2.5$\times$                \\
        Circa   & 66.4\%             & 2.6$\times$                \\
        AESPA   & 64.0\%             & 19.5$\times$                \\ 
        \Xhline{3\arrayrulewidth}
    \end{tabular}
    }
\end{table}


\end{document}